\documentclass[showpacs,superscriptaddress,preprintnumbers,amsmath,amssymb,twocolumn]{revtex4}

\usepackage{graphicx}% Include figure files
\usepackage{dcolumn}% Align table columns on decimal point
\usepackage{bm}% bold math
\usepackage{color}

\begin{document}

\title{Nonlinear Transport in a Two Dimensional Holographic Superconductor}

\author{Hua Bi Zeng}\email{zenghbi@gmail.com}
\affiliation{College of Physics Science and Technology, Yangzhou University, Jiangsu 225009, China}
\affiliation{Department of Physics, National Central University, Chungli 32001, Taiwan}
\affiliation{School of Mathematics and Physics, Bohai University, JinZhou 121000, China}

\author{Yu Tian}\email{ytian@ucas.ac.cn}
\affiliation{School of Physics, University of Chinese Academy of Sciences, Beijing 100049, China}
\affiliation{Shanghai Key Laboratory of High Temperature Superconductors, Shanghai 200444, China}

\author{Zhe Yong Fan}\email{brucenju@gmail.com}
\affiliation{COMP Centre of Excellence, Department of Applied Physics, Aalto University, Helsinki, Finland}

\author{Chiang-Mei Chen}\email{cmchen@phy.ncu.edu.tw}
\affiliation{Department of Physics, National Central University, Chungli 32001, Taiwan}

\begin{abstract}
The problem of nonlinear transport in a two dimensional superconductor with an applied oscillating electric field is solved by the holographic method. The complex conductivity can be computed from the dynamics of the current for both near- and non-equilibrium regimes. The limit of weak electric field corresponds to the near equilibrium superconducting regime, where the charge response is linear and the conductivity develops a gap determined by the condensate. A larger electric field drives the system into a superconducting non-equilibrium steady state, where the nonlinear conductivity is quadratic with respect to the electric field. Keeping increasing the amplitude of applied electric field results in a far-from-equilibrium non-superconducting steady state with a universal linear conductivity of one. In lower temperature regime we also find chaotic behavior of superconducting gap, which results in a non-monotonic field dependent nonlinear conductivity.
\end{abstract}


\maketitle
\pagebreak

\textit{Introduction}.---Charge transport of a system under a perturbative electric field $E \cos(\omega t)$ can be well understood by the linear response/Kubo formalism, since the properties of the system will be hardly affected by a small $E$. Nonlinear transport occurs naturally if we keep increasing the strength of the applied electric field, and the properties of the system depend in a singular way on $E$. However, understanding the nonlinear transport calls for a theory beyond the linear response theory, which is always a difficult task~\cite{Phillips2004, Green2005}, except for a system close to a quantum critical point where the appropriate non-equilibrium Green function and induced current can be calculated~\cite{Green2006, Green2008, Karch2011, Sonner2012, Green2013, Ribeiro2013}. Therefore, new insights and a general method are needed for studying a system with an arbitrary strength of field and away from a critical point.

The AdS/CFT correspondence~\cite{1, 2, 3, 4} states that the dynamics of a quantum many-body system can be obtained by solving the classical time evolution equation of its gravity dual, no matter the system is near or even far from equilibrium. For example, holography has been applied successfully to get insights of the superconducting gap dynamics for a long time evolution and far from equilibrium state in both spatial homogeneous configuration~\cite{Murata:2010dx, Bhaseen:2012gg, Gao:2012aw, Bai:2014tla, Li:2013fhw, Basu:2011ft, Basu:2012gg} and inhomogeneous configuration~\cite{Sonner:2014tca, Garcia-Garcia:2013rha, Adams2013, Du:2014lwa}. Many efforts have also been devoted to studying the superconducting equilibrium state phase transitions and charge transport properties in the linear response regime by following Ref.~\cite{Gubser:2008, HHH:2008}; for a review, see~\cite{Cai:2015cya}. The applications of holography to condensed matter are now known as AdS/CMT correspondence~\cite{Hartnoll:2009sz, McGreevy:2009xe, Sachdev:2010ch, Green:2013fqa}.

In condensed matter literature, previous works on the nonlinear charge transport in two dimensional superconductors mainly focused on the situation close to zero temperature quantum critical points between the superconducting and insulating states when a \textit{constant} electric field is applied~\cite{Phillips2004, Green2006}. Holography study of nonlinear conductivity focused on a non-superconducting steady current driven by a constant or oscillating electric field~\cite{Karch2011, Sonner2012, Horowitz:2013mia} and an $E$ and $\omega$ independent constant nonlinear conductivity was found. However, an investigation of the nonlinear complex conductivity corresponding to an \textit{oscillating} electric field in away from the equilibrium state is still lacking. An electric field like $E \cos(\omega t)$ will induce a time dependent pair breaking current in superconductor. The linear response theory can only address the regime of a very small $E$ in a static superconducting background. A large $E$ can induce a larger current which will drive the system out of equilibrium and suppress the superconducting gap, and eventually destroy superconductivity via a non-equilibrium phase transition~\cite{Li:2013fhw}.

Since holography provides an applicable method to easily compute the induced current by an external field, we extend the framework of holographic superconductor to study the real-time dynamics of current in both near-equilibrium and far-from-equilibrium regimes. With this method we are able to study nonlinear charge transport in non-equilibrium regime which is beyond the capability of the linear response theory. A specific $E$-dependent conductivity, $\sigma(\omega, E) \sim E^2$, appears in the far-infrared regime where the electric field can suppress the superconducting gap but is not strong enough to destroy it. By increasing $E$, the superconductivity will be destroyed at a critical value, and then the conductivity approaches to a universal value of one.

\textit{Model: Current dynamics and non-equilibrium phase transition.}---The action of $s$-wave holographic superconductor includes a $U(1)$ gauge field and a charged scalar
\begin{equation}
S = \int d^{4}x \sqrt{-g} \left( -\frac{1}{4} F_{\mu\nu} F^{\mu\nu} - |\nabla \Psi - i A \Psi|^2 - m^2 |\Psi|^2 \right),
\end{equation}
where we choose the mass parameter $m^2 = -2$ without loss of generality. The background is assumed to be the neutral $\textrm{AdS}_4$ planar black hole and its metric, in the retarded Eddington coordinates, reads
\begin{equation}
ds^2 = \frac{1}{z^2} \left( -f(z) dt^2 - 2 dt dz + dx^2 + dy^2 \right),
\end{equation}
where $f(z) = 1 - z^3$. The location of horizon is at $z = 1$, while $z = 0$ is the boundary where the field theory lives. According to the holographic dictionary, the gauge field in the bulk will source a conserved current $J$ on the boundary, while the scalar will source a scalar operator $O$ which breaks the $U(1)$ symmetry of the boundary field theory. Specifically, the asymptotical behaviors of the bulk fields on the boundary are,
\begin{equation}
\Psi = \Psi^{(1)} z + \Psi^{(2)} z^2, \quad A_\mu = a_\mu + b_\mu z.
\end{equation}
In the alternative quantization, the source term $\Psi^{(1)}$ is switched off to guarantee the appearance of a spontaneous symmetry broken phase by non-vanishing $\Psi^{(2)}$, and the expectation values of $O$ and $J$ are obtained by holography as the variation of renormalized bulk on-shell action with respect to the sources~\cite{Li:2013fhw}, i.e.,
\begin{equation} \label{OJ}
\langle O \rangle = \Psi^{(2)}, \quad J_\mu = - b_\mu + \partial_t a_\mu.
\end{equation}

\begin{figure}
\begin{center}
\includegraphics[trim=1.5cm 1.7cm 2cm 16.5cm, clip=true,scale=0.65]{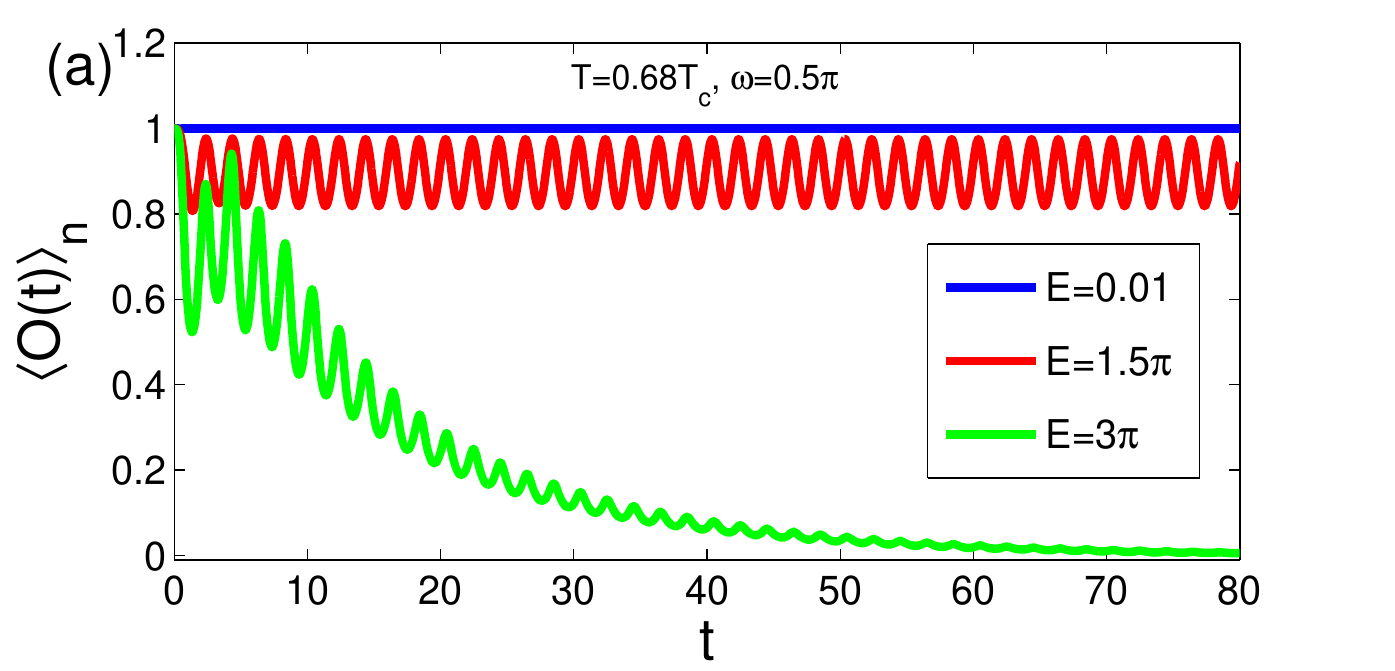}
\includegraphics[trim=1.5cm 1.7cm 2cm 15.7cm, clip=true,scale=0.65]{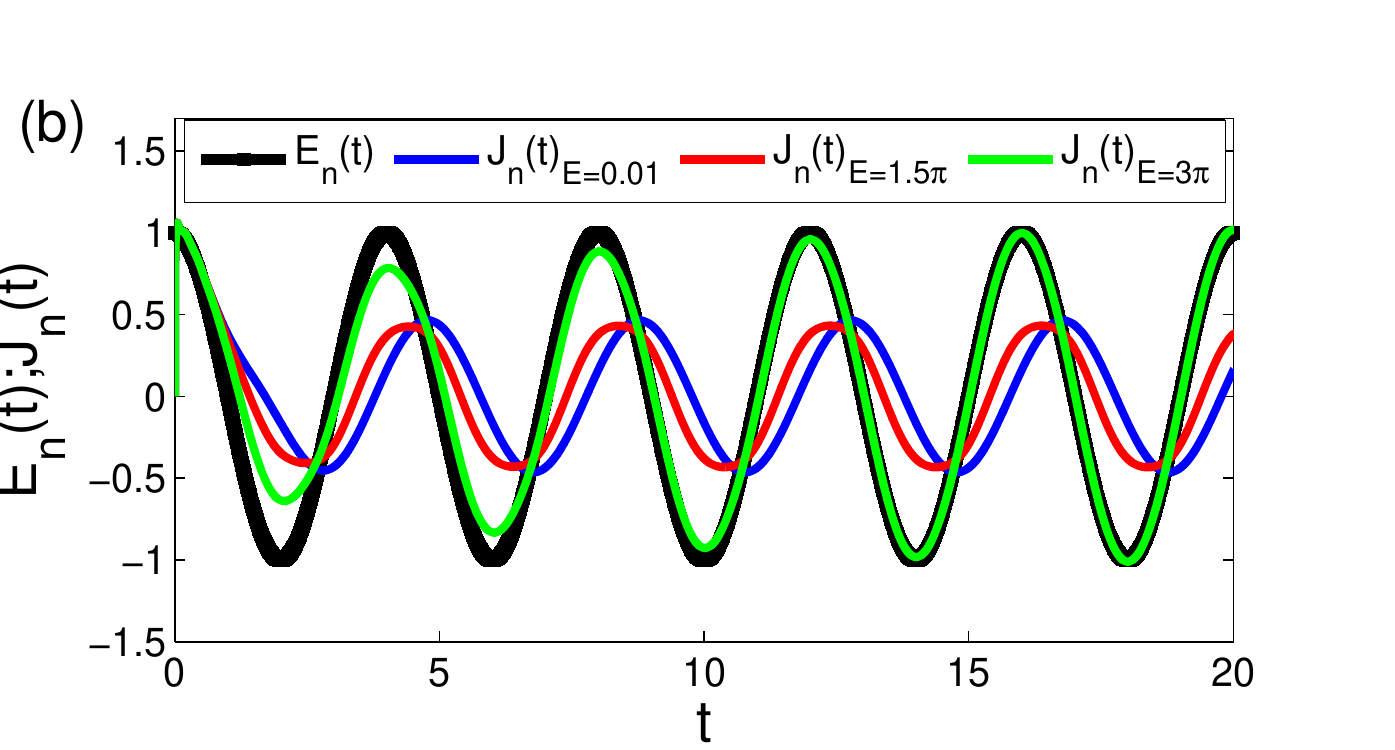}
\caption{(a): Superconducting gap dynamics $\langle O(t) \rangle_{n} = \frac{\langle O(t) \rangle}{\langle O(0) \rangle}$. (b): Current dynamics $J_{n}(t) = \frac{J(t)}{J(0)}$. $T = 0.68 T_c, \; \omega = 0.5 \pi$.} \label{fig1}
\end{center}
\end{figure}

The real-time dynamics of the superconductor is governed by the following time dependent equations of motion (EOMs):
\begin{eqnarray}
\partial_t \partial_z \Phi - i A_t \partial_z \Phi - \frac12 \Bigl[  f \partial_z^2 \Phi + f' \partial_z \Phi &&
\nonumber\\
+ i \partial_z A_t \Phi - z \Phi - A_x^2 \Phi \Bigr] &=& 0,
\\
\partial_t \partial_z A_t + 2 A_t |\Phi|^2 - i f (\Phi^* \partial_z \Phi - \Phi \partial_z \Phi^*) &&
\nonumber\\
+ i (\Phi^* \partial_t \Phi - \Phi \partial_t \Phi^*) &=& 0,
\\
\partial_t \partial_z A_x - \frac12 \left[ \partial_z ( f \partial_z A_x) - 2 A_x |\Phi|^2 \right] &=& 0,
\end{eqnarray}
combined with the ans\"atz of the following non-vanishing fields: $\Phi(t, z) = \Psi(t, z)/z, \, A_t(t, z), \, A_x(t, z)$. There is another constraint equation from the time component of Maxwell equations,
\begin{equation}
\partial_z (\partial_z A_t) - i (\Phi^* \partial_z \Phi - \Phi \partial_z \Phi^*) = 0.
\end{equation}

An AC electric field along the $x$ direction can be added by imposing the boundary condition at $z = 0$:
\begin{equation} \label{Ax}
A_x(t, z=0) = \frac{E \sin(\omega t)}{\omega}.
\end{equation}
Then the electric field is $E_x(t) = \partial_t A_x = E \cos(\omega t)$.  The initial condition at $t = 0$ is the static solution (time-independent) with a fixed chemical potential $A_t(z = 0) = \mu$ (the dimensionless temperature is defined by $T = 3/4 \pi \mu$ and the critical values are $\mu_c = 4.07$ and $T_c = 0.06$~\cite{HHH:2008}), which can be obtained by the spectral method. The EOMs will be solved by the fourth order Runge-Kutta method and the constraint equation is used to monitor the error of the solution by following Ref.~\cite{Li:2013fhw}.

Since the state density near the Fermi surface will oscillate driven by the oscillating current and the positive and negative current cause the same effect, the superconducting gap will oscillate with the frequency twice to the applied electric field~\cite{Li:2013fhw, Gurevich2014}, as seen from Fig.~1(a). In addition, a small-$E$ will hardly affect the superconducting gap leading the superconductor in a near equilibrium state. With increasing $E$, the system will be gradually driven out of equilibrium but still in a superconducting steady state. Eventually, for a sufficiently large $E$, a non-equilibrium phase transition from superconducting to normal state will be induced~\cite{Li:2013fhw}. At the transition point, the order parameter behaves as $(E_{c} - E)^{1/2}$, where the critical value $E_c$ depends on $\mu$ and $\omega$, and at a fixed $\mu$, $E_c \propto \omega$ for large frequency. 
%The mean field critical exponents is expected since the scaling and universality hold both in equilibrium and far from thermal equilibrium. When the fluctuation effect is suppressed in the large-$N$ limit (classical gravity theory dual), the non-equilibrium system with conserved order parameter exhibits the same critical behavior as the corresponding equilibrium system ~\cite{Schmittman1995, Odor2004, Feldman2005, Mitra2006}.

The current $J(t)$ is evaluated from Eq.~(\ref{OJ}), which oscillates with the same frequency of the driving electric field, as shown in Fig.~1(b). The amplitude of the oscillating current $J_\mathrm{max}$ reaches a constant value quickly and a time-independent conductivity is well defined when the current  develops a stable form of $J(t) = J_\mathrm{max} \sin(\omega t + \theta)$. For the value of $E = 3 \pi$, which is above the critical value $E_c$, the system enters the non-superconducting steady state, and the current is in phase with the electric field (green line), indicating that the conductivity is real. While for smaller electric fields $E = 0.01, 1.5 \pi$, the system is still in near equilibrium or out of equilibrium superconducting state, the current lags behind the electrical, which indicates that the conductivity is complex.

\begin{figure}
\begin{center}
\includegraphics[trim=2cm 2cm 0cm 17cm, clip=true,scale=0.65]{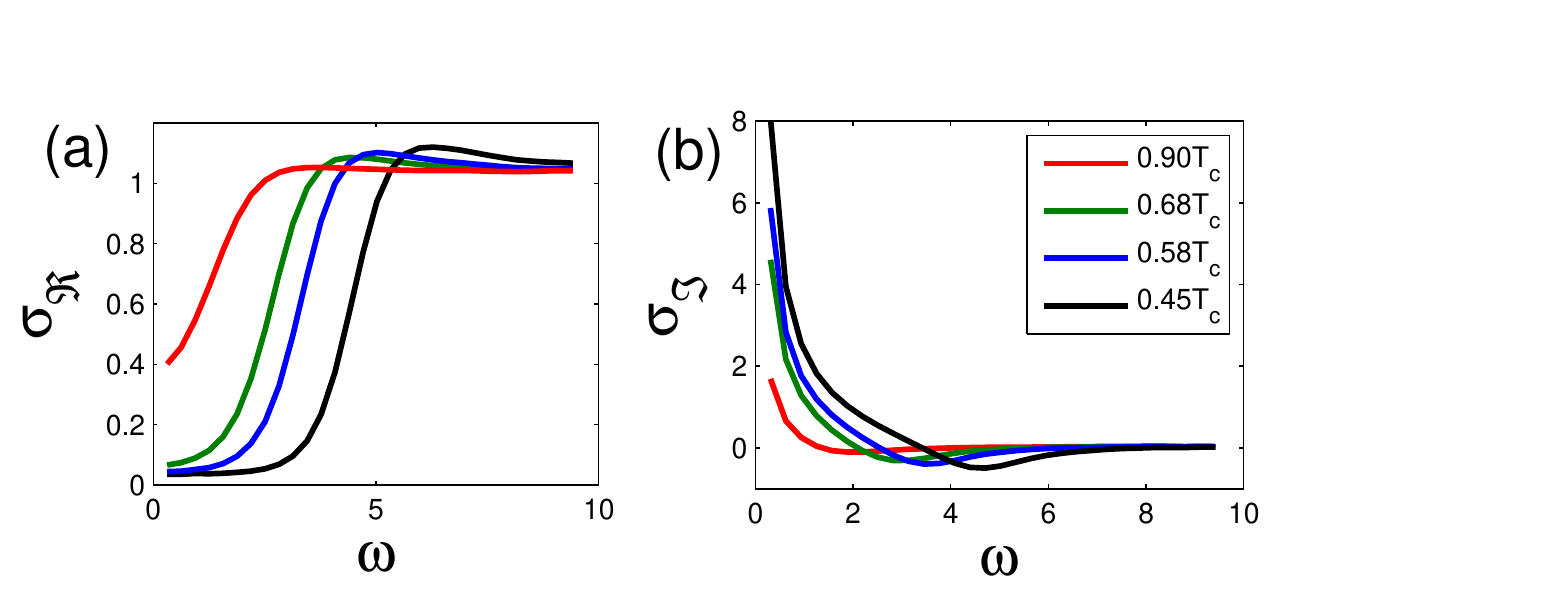}
\caption{Real part (a) and imaginary part (b) of the near equilibrium linear conductivity $\sigma(\omega)$ for different $T$ by fixing $E = 0.01 \ll \Delta$. The results are consistent with those in Ref.~\cite{HHH:2008}.} \label{fig2}
\end{center}
\end{figure}

If we have the real-time dynamics of the current, $\sigma(\omega)$ can be read off from the ``ratio'' of $J(t)$ to $E(t)$. This enables us to study the regime which is beyond the capability of the linear response theory. From $J(t) = \Re[ \sigma(\omega) E e^{i\omega t } ]$, and $E(t) = \Re(E e^{i \omega t }) = E \cos(\omega t )$, we have
\begin{equation} \label{Jt}
J(t)= %\Re [\sigma_{\Re}(\omega)+i \sigma_{\Im}(\omega))*(E_0 e^{i \omega t})] \\
E \left[ \sigma_{\Re}(\omega) \cos(\omega t ) - \sigma_{\Im}(\omega) \sin(\omega t ) \right].
\end{equation}
Therefore, the real and imaginary parts of the conductivity can be obtained by fitting the data of $J(t)$ and $E(t)$ after a steady state is achieved. 

\textit{Linear and non-linear conductivity.}---In the weak-field limit $E/\omega \ll \langle O \rangle$, the superconducting gap $\langle O \rangle$ remains a constant in time (blue line in Fig.~1(a)), and the gauge field $A_x$ basically can be treated as a perturbation with negligible back reaction to $A_t$ and $\Psi$~\cite{HHH:2008}. We use Eq.~({\ref{Jt}) to recompute $\sigma(\omega)$ in the weak-field limit including such back reaction. The results are shown in Fig.~2, which agree well with the linear response results in Ref.~\cite{HHH:2008}. The pole of $\sigma_{\Im}(\omega)$ at $\omega = 0$ indicates that the DC conductivity is infinite due to the Kramers-Kronig relation, which is a sign of superconducting state. The zero DC resistivity can also be observed by studying the $J(t)$ dynamics in a constant electric field by replacing Eq.~(\ref{Ax}) with $A_x(t, z=0) = E t$. We find that in this case $J(t)$ increases linearly in time initially and eventually approaches to the critical value at which the superconductivity will be destroyed no matter how small $E$ is. The linear increasing of current in time indicates a zero resistivity according to London's first equation $\partial_t J(t) \sim E(t)$. This gives us a hint that there is a critical $\omega$, below which the current will pass its maximum value and then the superconductivity will be destroyed. The minimal frequency we employ here is $\omega = 0.1 \pi$ which is larger than the critical value.

The superfluid density $n_s$ can be taken as the coefficient of the pole in the imaginary part of the complex conductivity according to
\begin{equation} \label{ns}
n_s \sim \omega \sigma_{\Im}(\omega), \qquad \omega \rightarrow 0,
\end{equation}
and the normal component is
\begin{equation} \label{nn}
n_n \sim \sigma_{\Re}, \qquad \omega \rightarrow 0.
\end{equation}
Moreover, the superconducting gap $\Delta$ can be fitted from the relation $n_n \sim e^{-\Delta/T}$, which gives $\Delta = \sqrt{\langle O \rangle}/2$~\cite{HHH:2008}. When $\omega \geq 2 \Delta$, the imaginary part falls to zero and the real part approaches to one~\cite{HHH:2008}. The normal state without condensation is found to have a constant conductivity $\sigma(\omega) = 1$ even for $\omega = 0$, which is a well known result in $\textrm{AdS}_4$~\cite{HHH:2008, Herzog:2007ij}.
%However, this constant conductivity is somehow unlike the actual conductor with dispersion, since we are working in the probe limit without any mechanism of momentum relaxation, and the DC conductivity should be infinite. One exception to this outcome is when there is zero net charge density, as is the case here. Under an applied electric field, the positive and negative charges move in opposite directions. They both contribute positively to the current, but cancel out in the momentum. So the current and momentum essentially decouple~\cite{Horowitz:2013mia}.

\begin{figure}
\begin{center}
\includegraphics[trim=1.5cm 1.5cm 2cm 16cm, clip=true,scale=0.55]{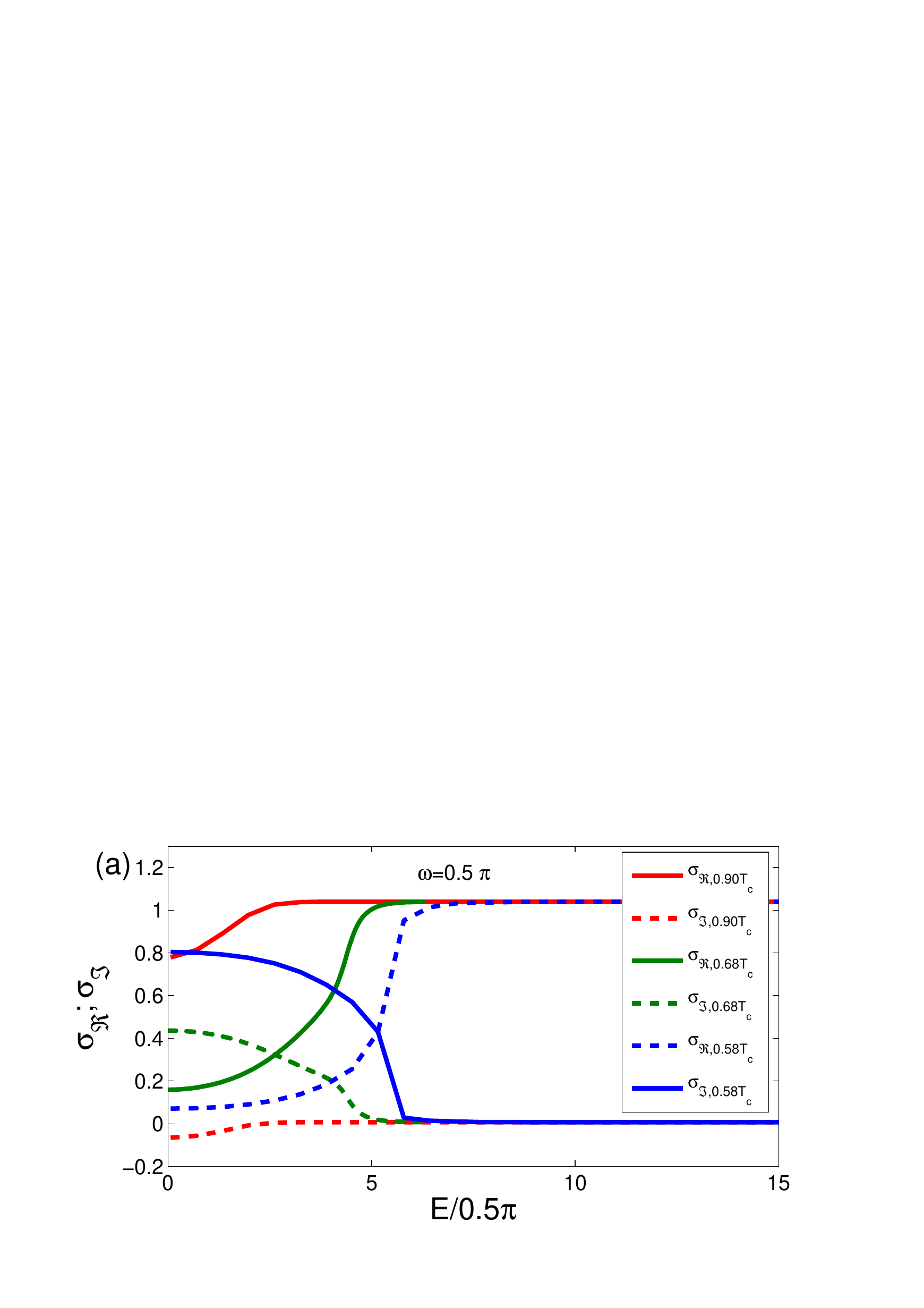}
\includegraphics[trim=1.5cm 1.5cm 2cm 16cm, clip=true,scale=0.55]{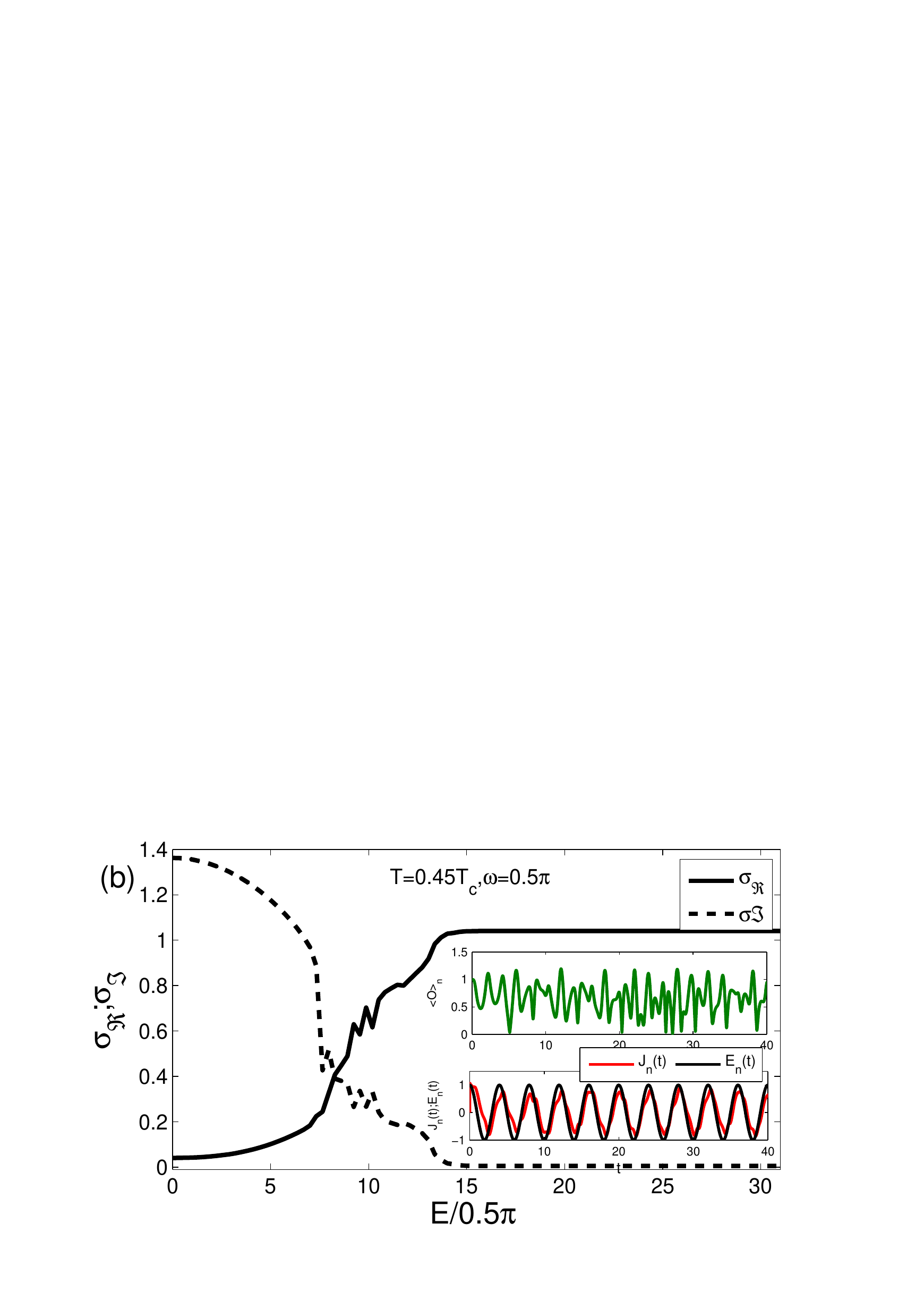}
\caption{(a) The behaviors of conductivity $\sigma(E)$ with $\omega = \pi/2$ at different temperatures. The critical value $E_c$ at which the conductivity approaches one increases by decreasing of temperature as a result of larger superconducting gap. (b) The non-monotonic dependence of $E$ appears at low temperatures (large $\mu$) due to the chaotic response to the electric field.} \label{fig3}
\end{center}
\end{figure}

Keeping increasing the electric field greater than $E = 0.01$ will drive the system away from the initial equilibrium state to another steady state (red, green lines in Fig.~1(a)), where the linear response theory is not able to give the correct conductivity. Fortunately, we still can analyze the real-time current dynamics via the holographic duality. It turns out that as the field exceeds a critical value, $E > E_c$, the superconductor will finally be driven into the normal conducting state. In this case, the non-equilibrium conductivity is $\sigma(\omega) = 1$, which can be clearly seen from the green line in Fig.~1(b) and the large-$E$ regime in Fig.~3. These results agree with previous holographic studies of current dynamics in steady states driven by an external electric field~\cite{Horowitz:2013mia}, where the dynamical metric driven by a constant electric field is included to accommodate the effect of Joule heating related to the linear growth of the black hole mass.
%We are working in the probe limit without considering the Joule heating, but Maxwell equation of $A_x$ is the same as Ref.~\cite{Horowitz:2013mia}. From the perspective of the dual field theory in the boundary, a possible physical reason for the constant conductivity is that the electric field will create massless particles via the Schwinger process which is described by the horizon~\cite{Sonner2012}, and the resulting increase in current can be balanced by the degradation due to particle scattering. An example of this balance is given in a field-theoretical model (also neglecting Joule heating) in Ref.~\cite{Green2005}.

From Fig.~3(a), the behavior of $\sigma(E)$ with fixed $\omega$ by increasing $E$ can be seen as follows: the real part will finally reach one, while the imaginary part will finally vanish. The $E \rightarrow 0$ results are the same as the equilibrium results in Fig.~2, since we start with an equilibrium superconducting state when $E$ is small. Actually, three different regimes can be identified: (1) a weak-field regime with linear equilibrium conductivity, which corresponds to a very small range of $E$ near $E = 0$, (2) an intermediate-field regime with non-equilibrium nonlinear conductivity, where $E$ is beyond the perturbative limit but smaller than the critical value $E_c$, and (3) a large-field regime with non-equilibrium but linear conductivity approaching to a constant $\sigma(\omega) = 1$.

The nonlinear behavior in the intermediate-field region is rather complicated. The nonlinear conductivity at the low frequency limit $\omega \rightarrow 0$ can be explained by a decreased superfluid density $n_s$ and an increase of normal density $n_n$. According to Eq.~(\ref{ns}) and Eq.~(\ref{nn}), the decrease of $n_s$ indicates that the imaginary part pole broadens with increasing $E$ and eventually disappears, while the increase of normal part $n_n$ results in an increase of the real part of the complex conductivity. For large frequency, since the conductivity for small $E$ is already about one, the conductivity basically does not vary as $E$ increases to nonlinear region. Thus, the significant nonlinear effect is more transparent for non-extremal frequency, and we choose $\omega = 0.5 \pi$ for our analysis. Remarkably, at lower temperature ($T = 0.45 T_c$) we have observed a non-smooth regime where both the superconducting gap and the induced current have chaotic dynamics; see Fig.~3(b). Moreover, at low $E$ regime, both the real and imaginary parts of conductivity with fixed $\omega$ for different temperatures, either close or away from $T_c$, can be fitted by a quadratic polynomial $\sigma(E) = a E^2 + b$ with two fitting parameters $a$ and $b$ (straight line part of the $d\sigma/dE$ curves in Fig.~4). The parameter $b$ is the value of the linear regime conductivity corresponding to weak field limit of $E$. Beyond this part, the system then has very diverse dynamics significantly depending on the control parameters.

\begin{figure}
\begin{center}
\includegraphics[trim=2.3cm 1.7cm 4cm 16cm, clip=true, scale=0.7]{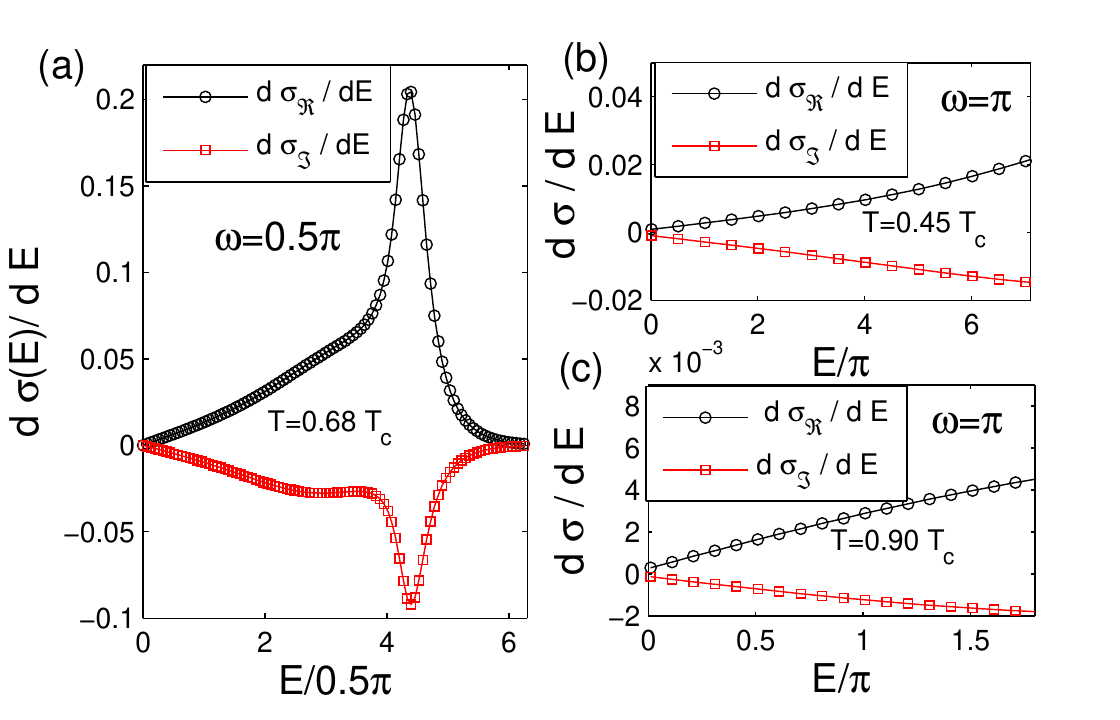}
\caption{Linear scaling of $\frac{d \sigma}{d E}= 2 a E $ in the low-field regime for different frequencies $\omega$ and temperatures, which confirms the universal $E^2$ scaling of the nonlinear conductivity when $E$ is small.} \label{fig4}
\end{center}
\end{figure}

\textit{Discussion.}---The $E$-dependent conductivity in the intermediate-field superconducting regime clearly cannot be obtained within the linear response theory. Interestingly, a similar $E^2$-scaling of conductivity has been found by Dalidovich and Philips in Ref.~\cite{Phillips2004}, where a two dimensional superconductor/insultor phase transition model with the same dynamic critical exponent $z = 2$ as the holographic superconductor~\cite{Maeda:2009wv, Jensen:2011af} was studied. However, they focus on the DC conductivity in the insulating side, but not in the superconducting side.
%We believe that the same $E^2$ scaling is just an incidence without any relation to our work.
In fact, the appearance of nonlinear AC conductivity we have found here can be explained as the suppression of the superconductivity due to the applied electric field $E \cos(\omega t)$. The normal part $n_n$ increases while the superconducting part $n_s$ decreases with the same scaling of $E^2$ respect to the electric field. 
%While the crossover to $E$ scaling at larger frequency is a combined effect of suppression of superconductivity and the quasi-particles excitation by the oscillation electric field. Furthermore, the nontrivial fitted curves in Fig.~4 in principle can be checked by experiments for relative small fields.

In order to check if these are universal results in any dimensions we should extend the discussion to a $d+1$ dimensional holographic superconductor dual to an $\textrm{AdS}_{d+2}$ gravity theory. One thing for sure is that in $\textrm{AdS}_3$ and $\textrm{AdS}_5$ the nonlinear conductivity in the non-superconducting steady state by large $E$ is not a constant any more~\cite{Horowitz:2013mia}.

\textit{Acknowledgments}---We thank Hai Qing Zhang for sharing his code and experiences. We thank Sean A. Hartnoll, Christopher P. Herzog, Shang-Yu Wu and Baruch Rosenstein for valuable comments. HBZ and ZYF are supported by the National Natural Science Foundation of China (under Grant No. 11205020 and No. 11404033). Y.T. is partially supported by NSFC with Grant No.11475179 and the Opening Project of Shanghai Key Laboratory of High Temperature Superconductors (14DZ2260700). CMC is supported by the Ministry of Science and Technology of Taiwan under the grant MOST 102-2112-M-008-015-MY3.

\end{document}